\documentclass{ws-ijmpe}

\newcommand{\beq}{\begin{equation}}
\newcommand{\eeq}{\end{equation}}
\newcommand{\beqa}{\begin{eqnarray}}
\newcommand{\eeqa}{\end{eqnarray}}
\newcommand{\bra}[1]{\mbox{$\langle #1|$}}
\newcommand{\ket}[1]{\mbox{$|#1\rangle$}}

\begin{document}

\markboth{E.M.\ Darwish}{Polarization observables of the $\gamma d\to\pi NN$ 
  reaction in the $\Delta$(1232)-resonance region}

\catchline{}{}{}{}{}
%
%%%%%%%%%%%%%%%%%%%%%%%%%%%%%%%%%%%%%%%%%%%%%%%%%%%%%%%%%%%%%%%%%%%%

\title{POLARIZATION OBSERVABLES OF THE $\gamma d\to\pi NN$ 
  REACTION IN THE $\Delta$(1232)-RESONANCE REGION}

\author{\footnotesize EED M.\ DARWISH\footnote{{\it E-mail address:}  
      eeddarwish@yahoo.com} }

\address{Physics Department, Faculty of Science, South Valley University, 
    Sohag 82524, Egypt}

\maketitle

\begin{history}
\received{(received date)}
\revised{(revised date)}
%\accepted{(Day Month Year)}
%\comby{(xxxxxxxxxx)}
\end{history}

\begin{abstract}
  Polarization observables of the three charge states of the pion for
  the $\gamma d\to\pi NN$ reaction with polarized photon beam and/or
  oriented deuteron target are evaluated over the whole
  $\Delta$(1232)-resonance region adopting a nonrelativistic model
  based on time-ordered perturbation theory. Results for the
  $\pi$-meson spectra, linear photon asymmetry, vector and tensor
  target asymmetries are presented. Particular attention is given, for
  the first time, to double polarization asymmetries for which we
  present results for $T_{20}^{\ell}$ and $T_{2\pm 2}^{\ell}$. We
  found that all other double polarization asymmetries of photon and
  deuteron target are vanished.
\end{abstract}

\section{Introduction} 
\label{sec1}
The study of pseudoscalar meson production in electromagnetic
reactions on light nuclei has become a very active field of research
in medium-energy nuclear physics with respect to the study of hadron
structure. For the following reasons the deuteron plays an outstanding
role besides the free nucleon. The first one is that the deuteron is
the simplest nucleus on whose structure we have abundant information
and a reliable theoretical understanding, i.e., the structure of the
deuteron is very well understood in comparison to heavier nuclei.
Furthermore, the small binding energy of nucleons in the deuteron,
which from the kinematical point of view provides the case of a nearly
free neutron target, allows one to compare the contributions of its
constituents to the electromagnetic and hadronic reactions to those
from free nucleons in order to estimate interaction effects.

Meson photo- and electroproduction on light nuclei is primarily
motivated by the following possibilities: (i) study of the elementary
neutron amplitude in the absence of a neutron target, (ii)
investigation of medium effects, i.e., possible changes of the
production operator in the presence of other nucleons, (iii) it
provides an interesting means to study nuclear structure, and (iv) it
gives information on pion production on off-shell nucleon, as well as
on the very important $\Delta N$-interaction in a nuclear medium.

The major reason for studying polarization phenomena lies in the fact
that only the use of polarization degrees of freedom allows one to
obtain complete information on all possible reaction matrix elements.
Without polarization, the cross section is given by the incoherent sum
of squares of the reaction matrix elements only. Thus, small
amplitudes are masked by the dominant ones.  On the other hand, small
amplitudes very often contain interesting information on subtle
dynamical effects. This is the place where polarization observables
enter, because such observables in general contain interference terms
of the various matrix elements in different ways. Thus, a small
amplitude may be considerably amplified by the interference with a
dominant matrix element.

Quasifree $\pi^-$ photoproduction on the deuteron via the $\gamma
d\to\pi^-pp$ reaction has been investigated within a diagrammatic
approach\cite{Log00}. In that work, the authors reported predictions
for the squared moduli of amplitudes $\mid
\hspace*{-0.13cm}T_{fi}\hspace*{-0.13cm}\mid^2$, analyzing powers
connected to beam polarization $T_{22,00}$, to target polarization
$T_{00,20}$, and to polarization of one of the final protons $P1_y$.
It has been shown, that final state interaction effects play a
noticeable role in the behaviour of these observables. In our previous
evaluation\cite{Dar03,Dar03+}, the energy dependence of the three
charge states of the pion for incoherent pion photoproduction on the
deuteron in the $\Delta$(1232)-resonance region has been investigated.
We have presented results for differential and total cross sections as
well as results for the beam-target spin asymmetry which determines
the Gerasimov-Drell-Hearn (GDH) sum rule.

Notwithstanding this continuing effort to study this process, the
wealth of information contained in it has not yet been fully
exploited. Since the $t$-matrix has 12 independent complex amplitudes,
one has to measure 23 independent observables, in principle, in order
to determine completely the $t$-matrix. Up to present times, only a
few observables have been measured and studied in detail, e.g.,
differential and total cross sections.

In view of the recent technical improvements, e.g., at MAMI in Mainz,
ELSA in Bonn and JLab in Newport News, for preparing polarized beams
and targets and for polarimeters for the polarization analysis of
ejected particles it appears timely to study in detail polarization
observables in pion production on the deuteron. The aim will be to see
what kind of information is buried in the various polarization
observables, in particular, what can be learned about the role of
subnuclear degrees of freedom like meson and isobar or even
quark-gluon degrees of freedom.

Most recently, we have investigated pion photoproduction on the
deuteron in the $\Delta$(1232)-resonance region with special emphasis
on single-spin asymmetries\cite{Dar04}. We have presented results for
the linear photon asymmetry $\Sigma$, vector target asymmetry
$T_{11}$, and tensor target asymmetries $T_{20}$, $T_{21}$, and
$T_{22}$ as functions of the emission pion angle for all the three
isospin channels of the reaction $\gamma d\to\pi NN$ with polarized
photon beam and/or oriented deuteron target. Our main goal in this
paper is to present predictions for the $\pi$-meson spectra,
single-spin asymmetries in a different kinematical situation, and, for
the first time, double polarization asymmetries of photon and deuteron
target. The difference between this paper and the former
one\cite{Dar04} lies in the fact that here more exclusive polarization
observables will be considered.

The paper is organized as follows. In Section \ref{sec2} we will
present the model for the elementary pion production amplitude which
will serve as an input for the reaction on the deuteron. Section
\ref{sec3} will introduce the general form of the differential cross
section for incoherent pion photoproduction on the deuteron. The
treatment of the $\gamma d\to\pi NN$ amplitude, based on time-ordered
perturbation theory, will be described in this section.  In Section
\ref{sec4} we will give the complete formal expressions of
polarization observables for the $\gamma d\to\pi NN$ reaction with
polarized photon beam and/or oriented deuteron target in terms of the
$t$-matrix elements. Details of the actual calculation and the results
will be presented and discussed in Section \ref{sec5}. Finally, we
close in Section \ref{sec6} with a summary and an outlook.
%%%%%%%%%%%%%%%%%%%%%%%%%%%%%%%%%%%%%%%%%%%%%%%%%%%%%%%%%%%%%%%%%%%%%%%%%%

\section{The $\gamma N\to\pi N$ Amplitude}
\label{sec2}
The electromagnetic production of pions on the free nucleon, including
photoproduction and electroproduction, has long been studied since the
pioneering work of Chew {\it et al.}\cite{Chew57}. As a result, an
enormous amount of knowledge has been accumulated. Recently,
theoretical interest in these reactions was revived by the new
generation of high-intensity and high duty-cycle electron
accelerators. With the developments of these new facilities, it is now
possible to obtain accurate data for meson electromagnetic production,
including spin-dependent observables. Extensive work during these more
than forty years (see for example\cite{Kroll54}$^-$\cite{Maid})
indicates that, below 500 MeV incident photon energy, the mechanisms
of the $\gamma N\rightarrow \pi N$ reaction are dominated by the Born
terms and the $\Delta(1232)$ excitation.

For the elementary pion photoproduction operator, we have taken in
this work, as in our previous work\cite{Dar03,Dar03+,Dar04}, the
effective Lagrangian model of Schmidt {\it et al.}\cite{ScA96}. This
model had been constructed to give a realistic description of the
$\Delta$(1232)-resonance region. It is given in an arbitrary frame of
reference and allows a well defined off-shell continuation as required
for studying pion production on nuclei. It consists of the standard
pseudovector Born terms and the contribution of the
$\Delta(1232)$-resonance. For further details with respect to the
elementary pion photoproduction operator we refer to the work of
Schmidt {\it et al.}\cite{ScA96}.
%%%%%%%%%%%%%%%%%%%%%%%%%%%%%%%%%%%%%%%%%%%%%%%%%%%%%%%%%%%%%%%%%%%%%%%%%%%

\section{Reaction on the Deuteron}
\label{sec3}
The formalism of incoherent pion photoproduction on the deuteron is
presented in details in our previous work\cite{Dar03}. We briefly
recall here the necessary notations and definitions.

The general expression of the cross section is given, using the
conventions of Bjorken and Drell\cite{BjD64}, by 
\beqa 
d\sigma &=&
(2\pi)^{-5}\delta^{4}\left( k+d-p_{1}-p_{2}-q\right)
\frac{1}{|\vec{v}_{\gamma}-\vec{v}_{d}|} \frac{1}{2}
\frac{d^{3}q}{2\omega_{\vec{q}}} \frac{d^{3}p_{1}}{E_{1}}
\frac{d^{3}p_{2}}{E_{2}}
\frac{M_{N}^{2}}{4\omega_{\gamma}E_{d}}\nonumber \\
& &\times~\frac{1}{6}\sum_{\alpha} |{\mathcal M}^{(t\mu)}_{s m
  m_{\gamma} m_d}|^{2} \, , 
\eeqa 
where $k=(\omega_\gamma,\vec k\,)$, $d=(E_d,\vec d\,)$,
$q=(\omega_q,\vec q\,)$, $p_1=(E_1,\vec p_1\,)$, and $p_2=(E_2,\vec
p_2\,)$ are the four-momenta of initial photon, deuteron, pion, and
two nucleons, respectively. Furthermore, $m_{\gamma}$ denotes the
photon polarization, $m_{d}$ the spin projection of the deuteron, $s$
and $m$ total spin and projection of the two outgoing nucleons,
respectively, $t$ their total isospin, $\mu$ the isospin projection of
the pion, and $\vec{v}_{\gamma}$ and $\vec{v}_{d}$ the velocities of
photon and deuteron, respectively. As a shorthand for the quantum
numbers we have introduced $\alpha=(s,m,t,m_{\gamma},m_d)$. The states
of all particles are covariantly normalized.  The reaction amplitude
is denoted by ${\mathcal M}^{(t\mu)}_{sm m_{\gamma}m_d}$.

For the evaluation we have chosen the laboratory frame where
$d^{\mu}=(M_d,\vec 0\,)$. As coordinate system a right-handed one is
taken with $z$-axis along the momentum $\vec k$ of the incoming photon
and $y$-axis along $\vec k\times\vec q$. Thus, the outgoing pion
defines the scattering plane. Another plane is defined by the momenta
of the outgoing nucleons which we will call the nucleon plane.

The fully exclusive differential cross section is given by
\beqa
\frac{d^5\sigma}{d\Omega_{p_{NN}} d\Omega_{\pi} dq} = \frac{\rho_{s}}{6}
\sum_{\alpha} |{\mathcal M}^{(t\mu)}_{sm m_{\gamma}m_d}|^{2}\,,
\label{fivefold}
\eeqa
where the phase space factor $\rho_{s}$ is expressed in terms of
relative and total momenta of the two final nucleons, $\vec p_{NN}$
and $\vec{P}_{NN}$, respectively, as 
\beqa
\label{rhos}
\rho_{s} &=& \frac{1}{(2\pi)^{5}}\frac{p_{NN}^{2}M_{N}^{2}}
{\left| E_{2} (p_{NN}+\frac{1}{2} P_{NN} 
\cos\theta_{Pp_{NN}}) + E_{1}
(p_{NN}-\frac{1}{2} P_{NN} \cos\theta_{Pp_{NN}}) \right| } \nonumber \\ 
& & \times~ \frac{q^{2}}{16\omega_{\gamma}M_{d}\omega_{q}} \, .
\eeqa
with $\theta_{Pp_{NN}}$ is the angle between $\vec{P}_{NN}$ and 
$\vec p_{NN}$.

The general form of the photoproduction transition matrix is given by
\beqa
{\mathcal M}^{(t\mu)}_{sm m_{\gamma}m_d}(\vec{k},\vec{q},\vec{p_1},\vec{p_2})
 &=&  ^{(-)}\bra{\vec{q}\,\mu,\vec{p_1}\vec{p_2}\,s\,m\,t-\mu}\epsilon_{\mu}
(m_{\gamma})J^{\mu}(0)\ket{\vec{d}\,m_d\,00}\, , 
\eeqa
where $J^{\mu}(0)$ denotes the current operator and
$\epsilon_{\mu}(m_{\gamma})$ the photon polarization vector. The outgoing 
$\pi NN$ scattering state is approximated in this work by the free $\pi NN$ 
plane wave, i.e.,
\beqa
\ket{\vec{q}\,\mu,\vec{p_1}\vec{p_2}\,s\,m\,t-\mu}^{(-)} & = &
\ket{\vec{q}\,\mu,\vec{p_1}\vec{p_2}\,s\,m\,t-\mu}\,.
\eeqa

The wave function of the final $NN$-state in a coupled spin-isospin basis 
which satisfies the symmetry rules with respect to a permutation of identical 
nucleons has the form
\beq
  |\vec{p}_{1},\vec{p}_{2},s m,t -\mu \rangle = 
  \frac{1}{\sqrt{2}}\left(
    |\vec{p}_{1}\rangle^{(1)}|\vec{p}_{2}\rangle^{(2)} - (-)^{s+t}
    |\vec{p}_{2}\rangle^{(1)}|\vec{p}_{1}\rangle^{(2)}\right)|s
  m\,,t -\mu\rangle\,,
\eeq
where the superscript indicates to which particle the ket refers.
In the case of charged pions, only the $t = 1$ channel contributes 
whereas for $\pi^{0}$ production both $t = 0$ and $t = 1$ channels
have to be taken into account. Then, one finds in the laboratory 
system for the matrix element the following expression
\beqa
\label{tmat_IA_lab}
{\mathcal M}_{sm m_{\gamma}m_d}^{(t\mu)} (\vec k,\vec q,\vec p_1,\vec
p_2) &=& \sqrt{2}\sum_{m^{\prime}}\langle s m,\,t -\mu|\,\Big( \langle
\vec{p}_{1}|t_{\gamma\pi}(\vec k,\vec q\,)|-\vec{p}_{2}\rangle
\tilde{\Psi}_{m^{\prime},m_{d}}(\vec{p}_{2})  \nonumber\\
& & \hspace{1cm} -(-)^{s+t}(\vec p_1 \leftrightarrow \vec p_2)
\Big)\,|1 m^{\prime},\,00\rangle.  
\eeqa 
where $t_{\gamma\pi}$ denotes the elementary production amplitude on
the nucleon and $\widetilde{\Psi}_{m,m_{d}}(\vec{p}\,)$ is given by
\beqa 
\widetilde{\Psi}_{m,m_{d}}(\vec{p}\,) &=&
(2\pi)^{\frac{3}{2}}\sqrt{2E_{d}} \sum_{L=0,2}\sum_{m_{L}}i^{L}\,C^{L
  1 1}_{m_{L} m m_{d}}\, u_{L}(p)Y_{Lm_{L}}(\hat{p}) \,, 
\eeqa
denoting with $C^{j_1 j_2 j}_{m_1 m_2 m}$ a Clebsch-Gordan
coefficient, $u_{L}(p)$ the radial deuteron wave function and
$Y_{Lm_{L}}(\hat{p})$ a spherical harmonics. The contribution to the
pion production amplitude in~(\ref{tmat_IA_lab}) is evaluated by
taking a realistic $NN$ potential model for the deuteron wave
function. For our calculations we have used the wave function of the
Paris potential\cite{La+81}.
%%%%%%%%%%%%%%%%%%%%%%%%%%%%%%%%%%%%%%%%%%%%%%%%%%%%%%%%%%%%%%%%%%%%%%%%%

\section{Theoretical Calculations of Polarization Observables}
\label{sec4}
The cross section for arbitrary polarized photons and initial deuterons can 
be computed for a given $\mathcal M$-matrix by applying the density matrix 
formalism similar to that given by Arenh\"ovel\cite{Aren88} for deuteron
photodisintegration. The most general expression for all possible polarization 
observables is given by 
\beqa
\mathcal O & = & \sum_{\alpha\alpha^{\prime}} \int d\Omega_{p_{NN}}~\rho_s~
\mathcal M^{(t^{\prime}\mu^{\prime})~\star}_{s^{\prime}m^{\prime},
m_{\gamma}^{\prime}m_d^{\prime}}~\vec{\Omega}_{s^{\prime}m^{\prime}sm}
~\mathcal M^{(t\mu)}_{sm,m_{\gamma}m_d}
~\rho^{\gamma}_{m_{\gamma}m_{\gamma}^{\prime}}~ \rho^{d}_{m_dm_{d}^{\prime}}\,,
\label{matho}
\eeqa
where $\rho^{\gamma}_{m_{\gamma}m_{\gamma}^{\prime}}$ and
$\rho^{d}_{m_dm_{d}^{\prime}}$ denote the density matrices of initial
photon polarization and deuteron orientation, respectively, 
$\vec{\Omega}_{s^{\prime}m^{\prime}sm}$ is an operator associated with
the observable, which acts in the two-nucleon spin space and $\rho_s$ 
is a phase space factor given in (\ref{rhos}). For further details we refer 
to Ref.\cite{Aren88,Rob74}.

As shown in Ref.\cite{Dar04,Aren88} all possible polarization observables for the 
pion photoproduction reaction on the deuteron can be expressed in terms of the 
quantities
\beqa
V_{IM}  &=& 
\frac{1}{2\sqrt{3}}~\sum_{m_d^{\prime}m_d}~\sum_{smt,m_{\gamma}}(-)^{1-m_d^{\prime}}  
\sqrt{2I+1} \left( \begin{array}{ccc}  1 & 1 & I \\
m_d & -m_d^{\prime} & -M \end{array} \right) \nonumber \\  
& &\times 
\int d\Omega_{p_{NN}} \rho_s  \mathcal M^{(t\mu)~\star}_{sm,m_{\gamma}m_d}\mathcal M^{(t\mu)}_{sm,m_{\gamma}m_d^{\prime}}\,,
\label{VIM}
\eeqa
and
\beqa
W_{IM} & = &
\frac{1}{2\sqrt{3}}~\sum_{m_d^{\prime}m_d}~\sum_{smt,m_{\gamma}}(-)^{1-m_d^{\prime}} 
\sqrt{2I+1} \left( \begin{array}{ccc}  1 & 1 & I \\
m_d & -m_d^{\prime} & -M \end{array} \right) \nonumber \\ 
& & \times 
\int d\Omega_{p_{NN}}\rho_s \mathcal M^{(t\mu)~\star}_{sm,m_{\gamma}m_d}\mathcal M^{(t\mu)}_{s-m,m_{\gamma}-m_d^{\prime}}\,,
\label{WIM}
\eeqa
where we use the convention of Edmonds\cite{Edm57} for the Wigner $3j$-symbols.

The unpolarized differential cross section is then given by
\beqa
\frac{d^3\sigma}{d\Omega_{\pi}dq}  &=&  V_{00}\,.
\eeqa
The photon asymmetry for linearly polarized photons is given by
\beqa
\Sigma ~\frac{d^3\sigma}{d\Omega_{\pi}dq}  &=&  - W_{00}\,.
\eeqa
The vector target asymmetry is given by
\beqa
T_{11}~\frac{d^3\sigma}{d\Omega_{\pi}dq}  &=&  2~\Im m V_{11}\,.
\label{T11}
\eeqa
The tensor target asymmetries are given by
\beqa
T_{2M} ~\frac{d^3\sigma}{d\Omega_{\pi}dq} & = & (2-\delta_{M0})~\Re e V_{2M}\,,~~
  (M=0,1,2)\,.
\eeqa
The photon and target double polarization asymmetries are given by \\
(i) Circular asymmetries
\beqa
T_{1M}^c ~\frac{d^3\sigma}{d\Omega_{\pi}dq}  &=&  (2-\delta_{M0})~\Re e
  V_{1M}\,,~~(M=0,1)\,,
\eeqa
\beqa
T_{2M}^c ~\frac{d^3\sigma}{d\Omega_{\pi}dq}  &=&  2~\Im m V_{2M}\,,~~(M=0,1,2)\,,
\eeqa
(ii) Longitudinal asymmetries
\beqa
T_{1M}^{\ell} ~\frac{d^3\sigma}{d\Omega_{\pi}dq}  &=&  i ~W_{1M}\,,~~(M=0,\pm 1)\,,
\eeqa
\beqa
T_{2M}^{\ell}~\frac{d^3\sigma}{d\Omega_{\pi}dq}  &=&  - W_{2M}\,,~~(M=0,\pm 1,\pm 2)\,.
\label{T2M}
\eeqa
%%%%%%%%%%%%%%%%%%%%%%%%%%%%%%%%%%%%%%%%%%%%%%%%%%%%%%%%%%%%%%%%%%%%%%%%%%

\section{Results and Discussion}
\label{sec5}
The discussion of our results is divided into three parts. First, we
will discuss the $\pi$-meson spectra as a function of the absolute
value of pion momentum $q$ at two different emission pion angles
$\theta_{\pi}=10^{\circ}$ and $120^{\circ}$ for photon energy at the
$\Delta$(1232)-resonance region, i.e. $\omega_{\gamma}^{\rm lab}=330$
MeV. In the second part, we will consider the single-spin asymmetries,
i.e., the linear photon asymmetry $\Sigma$, the vector target
asymmetry $T_{11}$, and the tensor target asymmetries $T_{20}$,
$T_{21}$, and $T_{22}$. In the last part, we will present and discuss
our results for the double polarization asymmetries. In all parts, we
will give calculations for the three isospin channels of the
$d(\gamma,\pi)NN$ reaction.
%%%%%%%%%%%%%%%%%%%%%%%%%%%%%%%%%%%%%%%%%%%%%%%%%%%%%%%%%%%%%%%%%%%%%%%%%%%

\subsection{The $\pi$-Meson Spectra}
\label{sec51}
\begin{figure}[th]
\centerline{\psfig{file=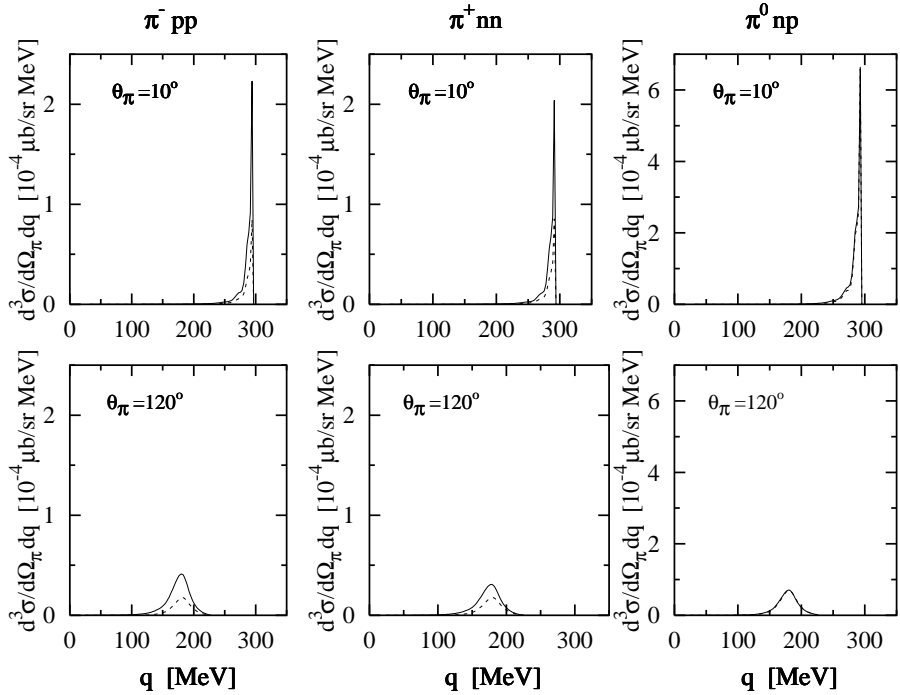,width=12cm}}
\caption{The $\pi$-meson spectra in the $d(\gamma,\pi)NN$ reaction as a 
function of the absolute value of pion momentum $q$ at a photon energy of 330 
MeV for two different values of emission pion angles $\theta_{\pi}$. The solid curves 
show the results of the full calculations while the dotted curves represent 
the results when only the $\Delta$(1232)-resonance is taken into account. 
The left, middle and right panels represent the results for 
$\gamma d\to\pi^-pp$, $\pi^+nn$ and $\pi^0np$, respectively.}  
\label{unpolcs1}
\end{figure}

We start the discussion with the $\pi$-meson spectra
$d^3\sigma/(d\Omega_{\pi}dq)$ depicted in Fig.~\ref{unpolcs1} as a
function of the absolute value of pion momentum $q$ at two different
values of emission pion angle $\theta_{\pi}$ for each isospin channel
of the $\gamma d\to\pi NN$ reaction for $\omega_{\gamma}^{lab}=330$
MeV.
One sees, that when the absolute value of pion momentum $q$ reaches
its maximum, the absolute value of the relative momentum $p_{NN}$ of
the two outgoing nucleons vanishes, and thus a narrow peak is appears
in the forward emission pion angles for charged as well as for neutral
pion photoproduction channels. In the lower part of
Fig.~\ref{unpolcs1} we see, that the unpolarized differential cross
section is small and the narrow peak which appears at forward emission
pion angles is disappears. The same effect appears in the coherent
process of charged pion photo- and electroproduction on the
deuteron\cite{Laget78,Kob87}, in deuteron
electrodisintegration\cite{Fab76} as well as in
$\eta$-photoproduction\cite{Fix97}. It is also clear that the maximum
value of $q$ (when $p_{NN}\to 0$) is decreases with increasing the
emission pion angle. In principle, the experimental observation of
this peak in the high $\pi$-momentum spectrum may serve as another
evidence for the understanding of the $\pi$-meson spectra.

In conclusion, one notes that the contributions from Born terms are
important for charged pion production channels but these are much less
important in the case of neutral pion production.
%%%%%%%%%%%%%%%%%%%%%%%%%%%%%%%%%%%%%%%%%%%%%%%%%%%%%%%%%%%%%%%%%%%%%%%%%%%

\subsection{Single-Spin Asymmetries}
\label{sec52}
\subsubsection{Linear Photon Asymmetry}
\label{sec521}
\begin{figure}[th]
\centerline{\psfig{file=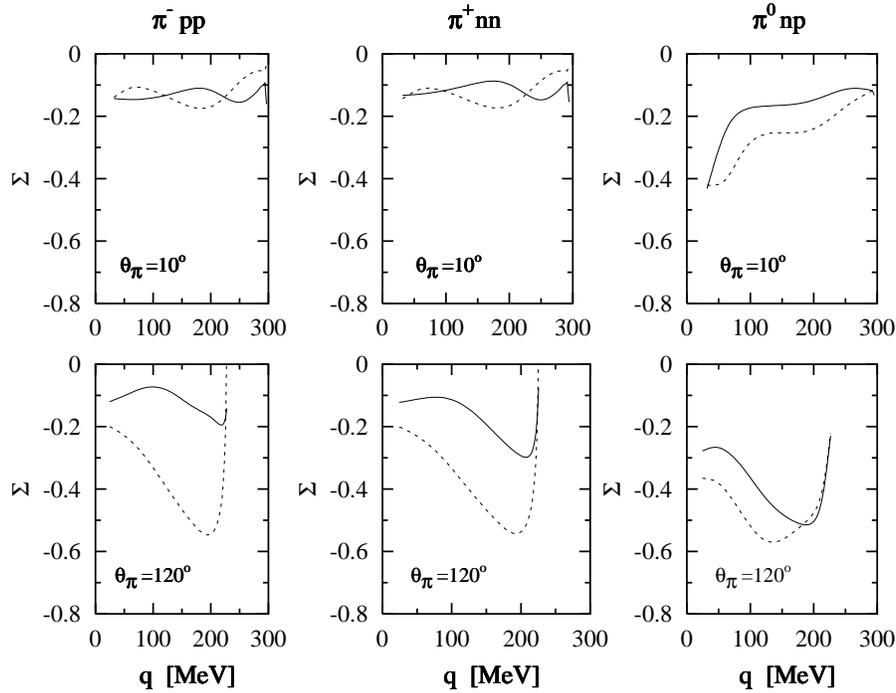,width=12cm}}
\caption{Linear photon asymmetry $\Sigma$ of 
  $d(\vec\gamma,\pi)NN$. Notation of the curves as in Fig.~\ref{unpolcs1}.}
\label{phasym1}
\end{figure}

Here we discuss our results for the photon asymmetry $\Sigma$ for
linearly polarized photons for all the different charge states of the
pion of $d(\vec\gamma,\pi)NN$. The $\gamma$-asymmetry for fixed pion
angles of $10^{\circ}$ and $120^{\circ}$ are plotted in
Fig.~\ref{phasym1} as a function of the absolute value of pion
momentum $q$ at $\omega_{\gamma}^{lab}=330$ MeV. The dotted curves
show the contribution of the $\Delta$(1232)-resonance alone in order
to clarify the importance of
the Born terms. We see that the photon asymmetry has always negative
values at forward and backward emission pion angles for charged as
well as for neutral pion channels. One notes qualitatively a similar
behaviour for charged pion channels whereas a totally different
behaviour is seen for the neutral pion channel.

It is also clear, that the contributions from Born terms are much
important, in particular at $q\simeq 200$ MeV which is very clear for
charged pion channels.  We observe that the interference of the Born
terms with the $\Delta$(1232)-resonance contribution causes
considerable changes in the photon asymmetry. Experimental
measurements as well as other theoretical predictions will give us
more valuable information on the photon asymmetry.
%%%%%%%%%%%%%%%%%%%%%%%%%%%%%%%%%%%%%%%%%%%%%%%%%%%%%%%%%%%%%%%%%%%%%%%%%%%

\subsubsection{Vector Target Asymmetry}
\label{sec522}
\begin{figure}[th]
\centerline{\psfig{file=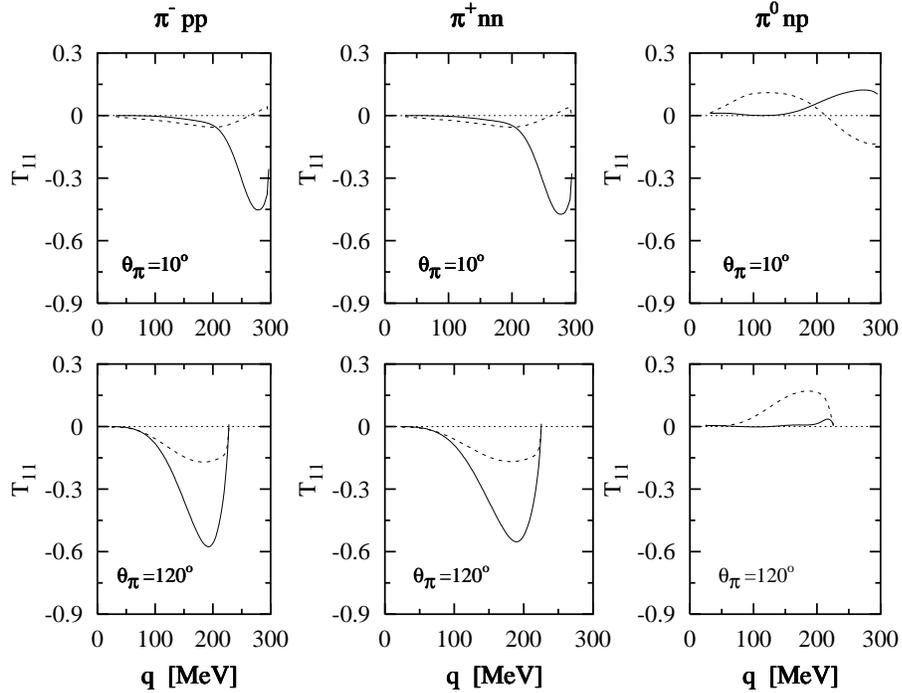,width=12cm}}
\caption{Vector target asymmetry $T_{11}$ of $\vec d(\gamma,\pi)NN$. 
  Notation of the curves as in Fig.~\ref{unpolcs1}.}
\label{vtasym1}
\end{figure}

Fig.~\ref{vtasym1} shows our results for the vector target asymmetry
$T_{11}$ as a function of $q$ at two different values of
$\theta_{\pi}$ for $\omega_{\gamma}^{lab}=330$ MeV. The asymmetry
$T_{11}$ clearly differs in size between charged and neutral pion
production channels, being even opposite in phase. For charged pion
production channels we see from the left and middle
panels of Fig.~\ref{vtasym1}, that the vector target asymmetry has
always negative values. At forward pion angles these values come
mainly from the Born terms since a small contribution from the
$\Delta$-resonance was found. At backward angles, the negative values
come from an interference of the Born terms with the
$\Delta$(1232)-resonance contribution since the $\Delta$-contribution
is large in this case.
\begin{figure}[th]
\centerline{\psfig{file=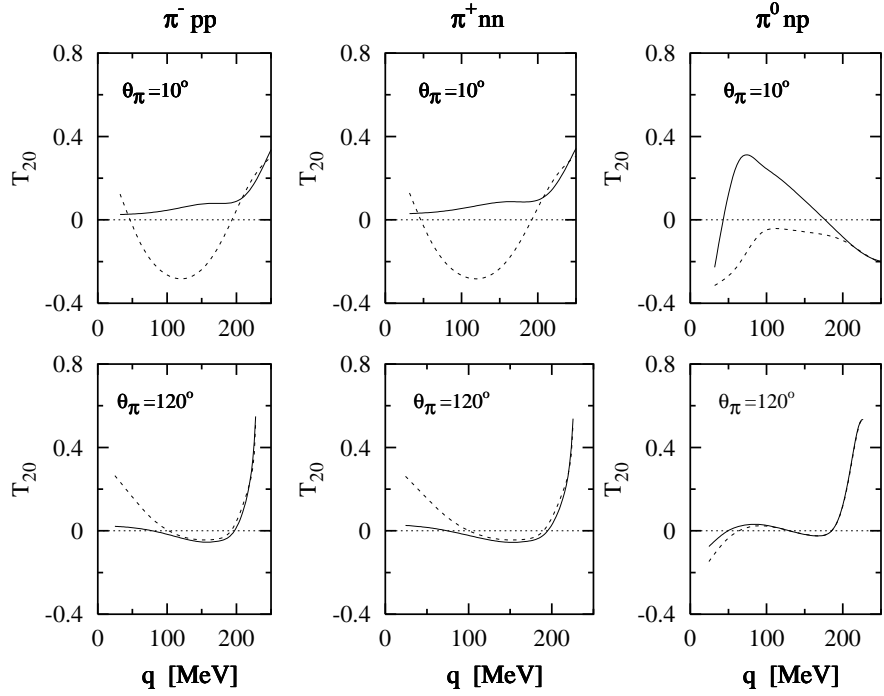,width=12cm}}
\caption{Tensor target asymmetry $T_{20}$ of $\vec d(\gamma,\pi)NN$. 
  Notation of the curves as in Fig.~\ref{unpolcs1}.}
\label{ttasym201}
\end{figure}

With respect to the neutral pion production channel, we see from the
solid curves of the right panel of Fig.~\ref{vtasym1}, that the vector
target asymmetry $T_{11}$ has a very small negative values at smaller
pion momentum and relatively large positive values at higher pion
momentum. It is interesting to point out the importance of the Born
terms in the charged pion production reactions in comparison to the
contribution of the $\Delta$(1232)-resonance. The sensitivity of
$T_{11}$ to the Born terms has also been discussed by Blaazer {\it et
  al.}\cite{Bla94} and Wilhelm and Arenh\"ovel\cite{Wil95} for the
coherent pion photoproduction reaction on the deuteron. The reason is
that $T_{11}$ depends on the relative phase of the matrix elements as
can be seen from (\ref{VIM}) and (\ref{T11}).  It would vanish for a
constant overall phase of the $t$-matrix, a case which is
approximately realized if only the $\Delta$(1232)-amplitude is
considered.
\begin{figure}[th]
\centerline{\psfig{file=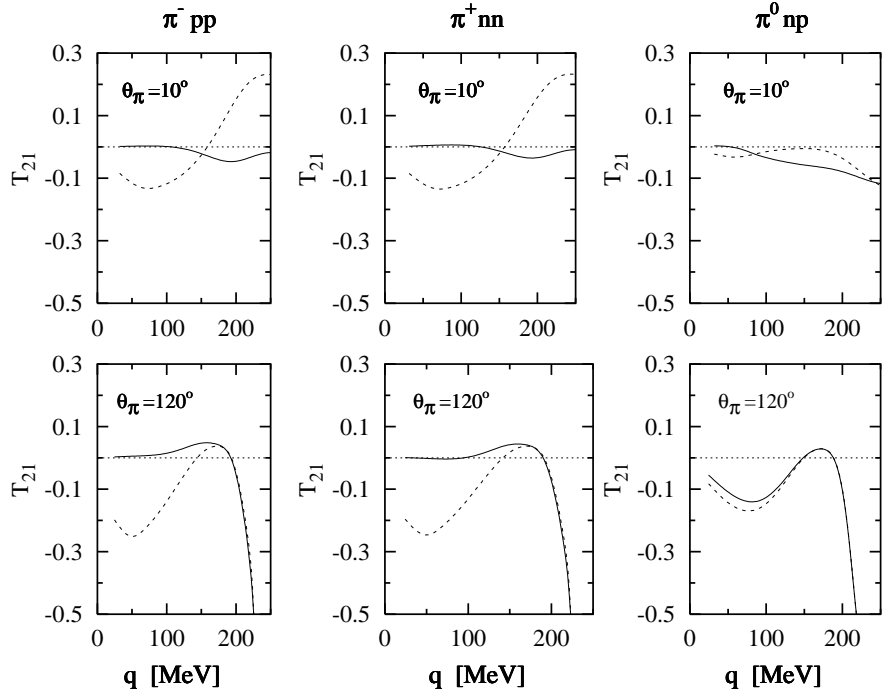,width=12cm}}
\caption{Tensor target asymmetry $T_{21}$ of $\vec d(\gamma,\pi)NN$. 
  Notation of the curves as in Fig.~\ref{unpolcs1}.}
\label{ttasym211}
\end{figure}
%
%%%%%%%%%%%%%%%%%%%%%%%%%%%%%%%%%%%%%%%%%%%%%%%%%%%%%%%%%%%%%%%%%%%%%%%%%%%

\subsubsection{Tensor Target Asymmetries}
\label{sec523}
Let us present and discuss now the results of the tensor target
asymmetries $T_{20}$, $T_{21}$, and $T_{22}$ for $\vec
d(\gamma,\pi)NN$.  We start from the tensor asymmetry $T_{20}$. For
$\gamma d\to\pi NN$ at forward and backward emission pion angles, the
asymmetry $T_{20}$ allows one to draw specific conclusions about
details of the reaction mechanism. Results for $T_{20}$ are plotted in
Fig.~\ref{ttasym201} at two different values of pion angles as a
function of $q$ for $\omega_{\gamma}^{lab}=330$ MeV. In general, one
notes again the importance of Born terms in the case of charged pion
production channels. In the case of neutral pion production channel
one sees, that the Born terms are important only at extreme forward
pion angles. One sees also that the contribution of Born terms is very
small for backward pion angles and higher pion momentum, but it is
relatively large for small pion momentum.

Fig.~\ref{ttasym211} shows our results for the tensor target asymmetry
$T_{21}$ as a function of $q$ for two fixed pion angles
$\theta_{\pi}=10^{\circ}$ and $120^{\circ}$ at $\omega_{\gamma}^{\rm
  lab}=330$ MeV.  One notices that the $T_{21}$ asymmetry is sensitive
to Born terms, in particular at forward pion angles. Also in this case
one notes the importance of Born terms in the case of charged pion
photoproduction reactions, in particular at smaller pion momentum. In
the case of $\pi^0$ channel one sees that the contribution of Born
terms is much less important at all angles.

\begin{figure}[th]
\centerline{\psfig{file=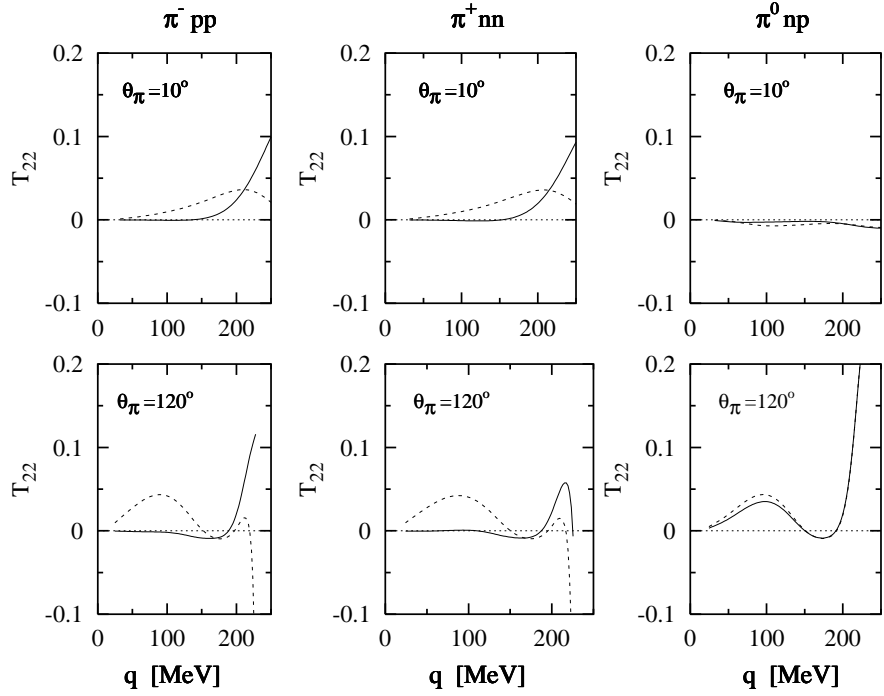,width=12cm}}
\caption{Tensor target asymmetry $T_{22}$ of $\vec d(\gamma,\pi)NN$. 
  Notation of the curves as in Fig.~\ref{unpolcs1}.}
\label{ttasym221}
\end{figure}

In Fig.~\ref{ttasym221} we depict our results for the tensor target
asymmetry $T_{22}$ as a function of $q$. We have used here the same
two values of pion angle $\theta_{\pi}$ as in the previous figures.
Like the results of Figs.~\ref{ttasym201} and \ref{ttasym211}, the
$T_{22}$ asymmetry is sensitive to the values of pion angle
$\theta_{\pi}$. We notice that the $T_{22}$ asymmetry changes
dramatically if only the $\Delta$-contribution is taken into account.
%
%%%%%%%%%%%%%%%%%%%%%%%%%%%%%%%%%%%%%%%%%%%%%%%%%%%%%%%%%%%%%%%%%%%%%%%%%%

\subsection{Double Polarization Asymmetries}
\label{sec53}
In this subsection we present and discuss our results, obtained with
the deuteron wave function of the Paris potential\cite{La+81} and the
pion production operator on the free nucleon of Schmidt {\it et
  al.}\cite{ScA96}, for the circular and longitudinal double
polarization asymmetries.  The results of this section are presented
as functions of pion angle $\theta_{\pi}$ at different values of
photon lab-energy, i.e., we integrated over $q$ from $0$ to $q_{max}$
in Eqs.~(\ref{matho},\ref{VIM},\ref{WIM}).

First of all, we would like to mention here that all the circular
double polarization asymmetries are vanished in the region of the
$\Delta$(1232)-resonance. With respect to the longitudinal double
polarization asymmetries, we found that only the $T_{20}^{\ell}$ and
$T_{2\pm 2}^{\ell}$ asymmetries are not vanished.  All other
longitudinal double polarization asymmetries are vanished in the
region of our interest.  However, it is difficult to measure these
asymmetries in the $\Delta$(1232)-resonance region. We found also that
the values of the $T_{2+2}^{\ell}$ asymmetry are identical with the
values of $T_{2-2}^{\ell}$. Therefore, in the following we present and
discuss our results for only $T_{20}^{\ell}$ and $T_{2+2}^{\ell}$.

\subsubsection{The Double Polarization Asymmetry $T_{20}^{\ell}$}
\label{sec531}
\begin{figure}[th]
\centerline{\psfig{file=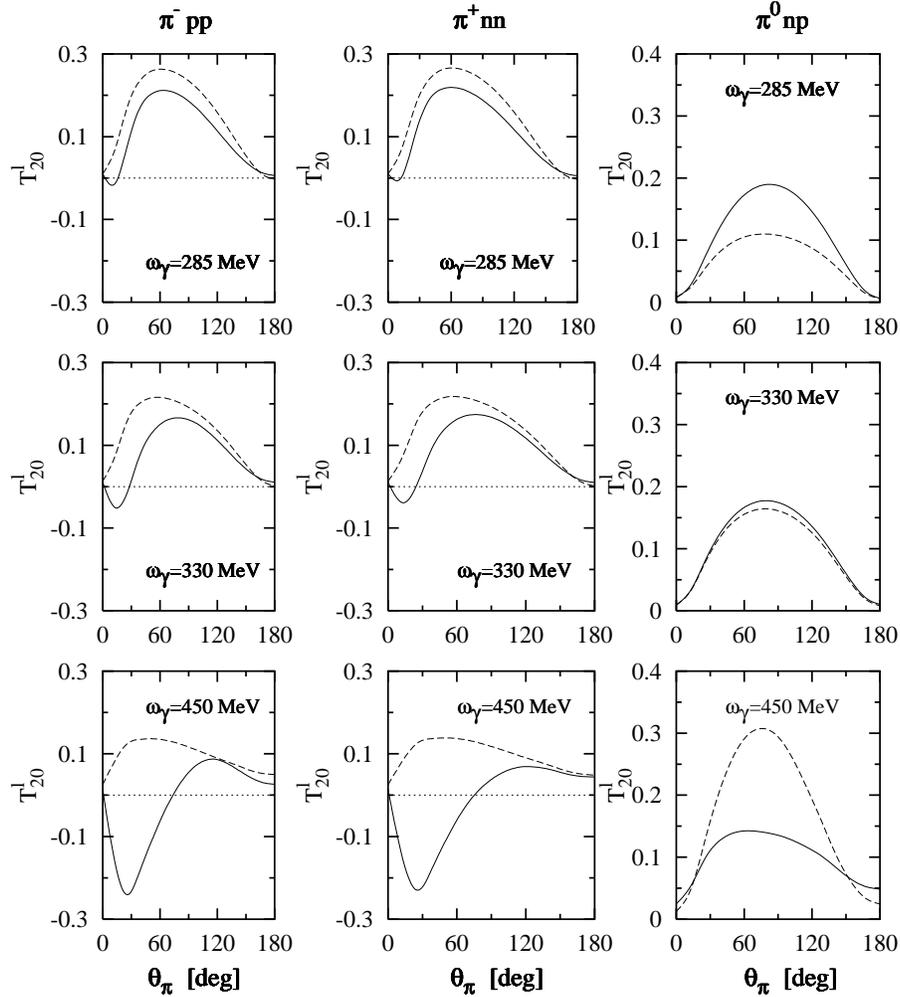,width=12cm}}
\caption{The double polarization asymmetry $T_{20}^{\ell}$ of 
  $\vec d(\vec\gamma,\pi)NN$ as a function of emission pion angle $\theta_{\pi}$ 
  at three different values of photon lab-energies. Notation of the curves as 
  in Fig.~\ref{unpolcs1}.}
\label{dsasymt20l}
\end{figure}

Fig.~\ref{dsasymt20l} shows our results for the longitudinal double
polarization asymmetry $T_{20}^{\ell}$ (see Eqs.~(\ref{WIM}) and
(\ref{T2M}) for its definition) as a function of emission pion angle
$\theta_{\pi}$ for fixed values of photon lab-energies of
$\omega_{\gamma}^{lab}=$285, 330, and 450 MeV for all the three charge
states of the pion for the reaction $\vec\gamma \vec d\to\pi NN$.  For
neutral pion production (see the right panels of
Fig.~\ref{dsasymt20l}), we see that $T_{20}^{\ell}$ has always
positive values for all photon lab-energies and emission pion angles.
For energies below and above
the $\Delta$(1232)-resonance, the double polarization asymmetry
$T_{20}^{\ell}$ shows sensitivity on the Born terms, in particular for
angles between $60^{\circ}$ and $120^{\circ}$. Furthermore, it is
apparent that at extreme forward and backward emission pion angles the
asymmetry $T_{20}^{\ell}$ is very small in comparison to other angles.
At $\omega_{\gamma}^{lab}=$450 MeV the dominant contribution comes
from the resonance term.

For the calculations of charged pion production channels (see the left
and middle panels of Fig.~\ref{dsasymt20l}), we see that
$T_{20}^{\ell}$ has negative values at forward pion angles which is
not the case at backward angles. It is also noticeable that the
contributions of Born terms are large, in particular at energies above
the resonance region. At extreme backward angles, we see that
$T_{20}^{\ell}$ has small positive values.

\subsubsection{The Double Polarization Asymmetry $T_{2+2}^{\ell}$}
\label{sec532}
\begin{figure}[th]
\centerline{\psfig{file=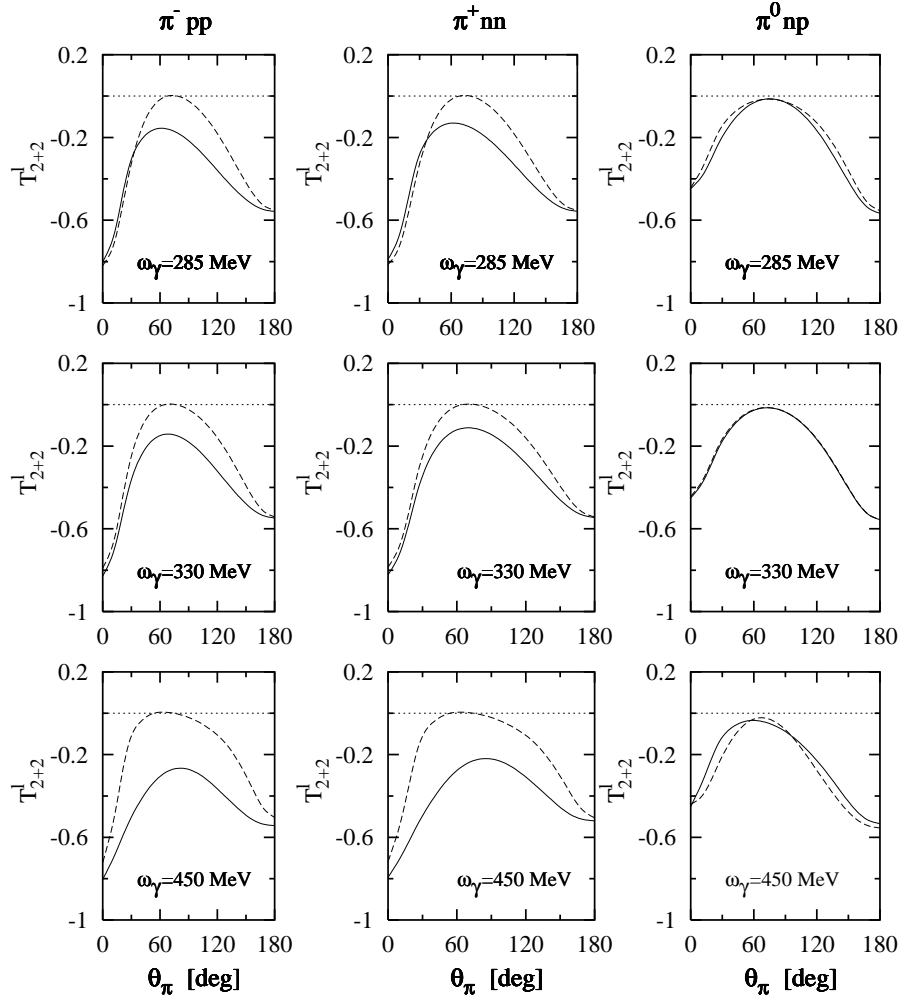,width=12cm}}
\caption{The double polarization asymmetry $T_{2+2}^{\ell}$ of $\vec d(\vec\gamma,\pi)NN$ 
  as a function of emission pion angle $\theta_{\pi}$ at three different 
  values of photon lab-energies. Notation of the curves as in 
  Fig.~\ref{unpolcs1}.}
\label{dsasymt2p2l}
\end{figure}

First of all, we would like to point out that the values of
$T_{2+2}^{\ell}$ asymmetry are identical with the values of
$T_{2-2}^{\ell}$ (see Eqs.~(\ref{WIM}) and (\ref{T2M}) for their
definition). Therefore, the results of the $T_{2-2}^{\ell}$ asymmetry
are not presented here. Our results for the longitudinal double
polarization asymmetry $T_{2+2}^{\ell}$ are plotted in
Fig.~\ref{dsasymt2p2l} for the photon lab-energies
$\omega_{\gamma}^{lab}=$285, 330, and 450 MeV as a function of
emission pion angle for all the three charge states of the pion for
the reaction $\vec\gamma \vec d\to\pi NN$. The solid curves show the
results of the full calculations while the dotted curves represent the
results when only the $\Delta$(1232)-resonance is taken into account.

In general, one readily notes that the longitudinal asymmetry
$T_{2+2}^{\ell}$ has negative values at all photon energies and pion
angles. For neutral pion production channel (see the right panels of
Fig.~\ref{dsasymt2p2l}), we noticed that the contributions of Born
terms are negligible, even at energies below and above the resonance.
For charged pion production channels (see the left and middle panels
of Fig.~\ref{dsasymt2p2l}), it is apparent that the contributions of
Born terms reduce the $T_{2+2}^{\ell}$ asymmetry. It is very
interesting to examine these spin observables experimentally.
%%%%%%%%%%%%%%%%%%%%%%%%%%%%%%%%%%%%%%%%%%%%%%%%%%%%%%%%%%%%%%%%%%%%%%%%%%

\section{Conclusions and Outlook}
\label{sec6}
In this paper we have presented results for $\pi$-meson spectra,
single and double polarization asymmetries for incoherent single pion
photoproduction reaction on the deuteron in the
$\Delta$(1232)-resonance region. The $\gamma d\to\pi NN$ scattering
amplitude is given as a linear combination of the on-shell matrix
elements of pion photoproduction on the two nucleons. For the
elementary pion photoproduction operator an effective Lagrangian model
is used which is based on time-ordered perturbation theory and
describes well the elementary $\gamma N\to\pi N$ reaction.

Predictions for unpolarized differential cross section
$d^3\sigma/d\Omega_{\pi}dq$, single polarization asymmetries (linear
photon asymmetry $\Sigma$, vector target asymmetry $T_{11}$ and tensor
target asymmetries $T_{20}$, $T_{21}$, and $T_{22}$) as well as for
double polarization asymmetries (circular and longitudinal double
polarization asymmetries for photon and deuteron target) are given. We
found that all the circular double polarization asymmetries are
vanished. Concerning the longitudinal double polarization asymmetries,
we found that only $T_{20}^{\ell}$ and $T_{2\pm 2}^{\ell}$ are not
vanished. All other double polarization asymmetries are vanished in
the region of our interest. As already noticed in the discussion
above, we found also that interference of Born terms and the
$\Delta$(1232)-contribution plays a significant role. Unfortunately,
there are no experimental data available to be compared to the
observables we computed.

We would like to conclude that the results presented here for
polarization observables in the $d(\gamma,\pi)NN$ reaction in the
$\Delta$-resonance region can be used as a basis for the simulation of
the behaviour of polarization observables and for an optimal planning
of new polarization experiments of this reaction. It would be very
interesting to examine our predictions experimentally.

Finally, we would like to point out that future improvements of the
present approach should include further investigations including final
state interactions. This approach is necessary for the problem at hand
since polarization observables are sensitive to dynamical effects.  As
future refinements we consider also the use of a more sophisticated
elementary production operator, which will allow one to extend the
present approach to higher energies, and the role of irreducible
two-body contributions to the electromagnetic pion production
operator.

\section*{Acknowledgements}
I would like to thank H.\ Arenh\"ovel as well as many 
scientists of the Institut f\"ur Kernphysik of the J.\ 
Gutenberg-Universit\"at, Mainz for fruitful discussions.

\end{document}